\documentclass{JAC2003}

\usepackage{graphicx}
\usepackage{booktabs}

\begin{document}
\title{
SECOND-ORDER ACHROMATS WITH ARBITRARY LINEAR TRANSFER MATRICES
\vspace{-0.5cm}}
\author{V.Balandin\thanks{vladimir.balandin@desy.de}, 
R.Brinkmann, W.Decking, N.Golubeva \\
DESY, Hamburg, Germany}

\maketitle

\begin{abstract}
In this article we consider a system where a bend magnet block arranged 
in an achromat-like fashion is followed by a straight drift-quadrupole cell
which is not a pure drift space.
We formulate the necessary and sufficient conditions for this system 
to be a second-order achromat and show that
it can be achieved using six, four or even only two sextupole families.
\end{abstract}

\vspace{-0.3cm}
\section{INTRODUCTION}

\vspace{-0.1cm}
As a second-order achromat we will understand a particle transport system
whose linear transfer matrix is dispersion free and whose transfer map does 
not have transverse second-order aberrations.
The first practical solution for the second-order achromat was presented
at the end of 1970s in the paper ~\cite{Brown}, where the theory of achromats 
based on repetitive symmetry was developed,
and quickly becomes part of many accelerator designs.
Unfortunately, the overall transfer matrix of this achromat is always equal 
to the identity matrix (except, possibly, for the $r_{56}$ element)
and variety of transfer matrices of all other known
second-order achromats is also very limited.
The most natural way to satisfy a need for a second-order achromat with an 
arbitrary linear transfer matrix, as it seems at first sight,
is to take a bend magnet system arranged in an achromat-like fashion with the total
transfer matrix equal to the identity matrix, attach a drift-quadrupole
block with the desired linear transfer matrix and then adjust the sextupoles 
installed in the first part in such a way that all 
transverse second-order aberrations of the
total system are canceled.
In this paper, using the group-theoretical point of view for the design of
magnetic optical achromats developed in ~\cite{AchromIPAC10},
we formulate the necessary and sufficient conditions for this system 
to be a second-order achromat and show that it can be 
achieved using six, four or two sextupole families.
We also show that if one uses less than six sextupole families, then
the linear transport in the achromat-like part cannot be designed independently from
the properties of the attached straight drift-quadrupole cell.

\vspace{-0.2cm}
\section{DYNAMICAL VARIABLES AND MAPS}

\vspace{-0.1cm}
We will consider the beam dynamics in a mid-plane symmetric
magnetostatic system and will use a complete set of symplectic variables
$\mbox{\boldmath $z$} = (x, p_x, y, p_y, \sigma, \varepsilon)^{\top}$
as particle coordinates. 
In this set the variables $\hat{\mbox{\boldmath $z$}} = (x, p_x, y, p_y)^{\top}$
describe the transverse particle motion and 
the variables $\sigma$ and $\varepsilon$ characterize
the longitudinal dynamics ~\cite{AchromIPAC10, ApochrIPAC10}.
We will represent particle transport from one longitudinal location to another
by a symplectic map and 
we will assume that for arbitrary two longitudinal positions 
the point $\mbox{\boldmath $z$} = \mbox{\boldmath $0$}$ is the fixed
point and that the corresponding map can be Taylor 
expanded in its neighborhood.
We will use that up to any predefined order $m$ the aberrations 
of a map $\,{\cal M}\,$ can be represented through a Lie factorization as 

\vspace{-0.1cm}
\noindent
\begin{eqnarray}
:{\cal M}: \,=_m\,
\exp(:{\cal F}_{m + 1} + \ldots + {\cal F}_3:) :M: ,
\label{IFB_3}
\end{eqnarray}

\vspace{-0.1cm}
\noindent
where each of the functions ${\cal F}_k$
is a homogeneous polynomial of degree $k$ in the variables $\mbox{\boldmath $z$}$
and the symbol $=_m$ denotes equality up to order $m$ (inclusive)
when maps on both sides of (\ref{IFB_3})
are applied to the phase space vector $\mbox{\boldmath $z$}$.
We will also use that for the map ${\cal M}$ of a magnetostatic system which is 
symmetric about the horizontal midplane $\,y = 0\,$ 
all polynomials ${\cal F}_k$  in (\ref{IFB_3}) 
do not depend on the variable $\sigma$ and are even functions of the
variables $y$ and $p_y$.

\vspace{-0.2cm}
\section{SECOND-ORDER ABERRATIONS OF REPETITIVE SYSTEM ARRANGED IN ACHROMAT-LIKE FASHION}

In this section we will consider a system constructed by a repetition of $n$ identical 
cells ($n > 1$) with the cell map ${\cal M}_c$ given by the following Lie factorization

\vspace{-0.1cm}
\noindent
\begin{eqnarray}
:{\cal M}_c: \,=_2\,
\exp(:{\cal F}_3^c(\mbox{\boldmath $z$}):) :M_c:.
\label{TWO_C_1}
\end{eqnarray}

\vspace{-0.1cm}
Let two by two symplectic matrices $M_{cx}$ and $M_{cy}$ be the horizontal and 
vertical focusing blocks of the six by six cell transfer matrix $M_c = (r_{mk}^c)$
and let us define the four by four cell transverse focusing matrix $\hat{M}_c$ as follows

\vspace{-0.1cm}
\noindent
\begin{eqnarray}
\hat{M}_c \,=\, \mbox{diag}\left(M_{cx},\,M_{cy}\right).
\label{TWO_C_1_HAT}
\end{eqnarray}

\vspace{-0.1cm}
We will say that a repetitive $n$-cell system is arranged in achromat-like fashion
if its linear transfer matrix $M_c^n$ is dispersion free and if the cell transverse 
focusing matrix $\hat{M}_c$ generates a cyclic group of order $n$, which 
means that 

\vspace{-0.1cm}
\noindent
\begin{eqnarray}
\hat{M}_c^n = I
\;\;\;
\mbox{and}
\;\;\;
\hat{M}_c^m \neq I
\;\;\;
\mbox{for}
\;\;\;
m = 1, \ldots, n-1. 
\label{TWO_C_2_HAT}
\end{eqnarray}

\vspace{-0.6cm}
\subsection{Dispersion Decomposition of the Cell Matrix}

\vspace{-0.1cm}
If the $n$-cell system is arranged in the achromat-like fashion, then
its linear transfer matrix $M_c^n$ is equal to the identity matrix (except,
possibly, for the $r_{56}$ element) and, as a consequence of this, the equations

\vspace{-0.2cm}
\noindent
\begin{eqnarray}
M_{cx}^n \,=\, I,
\label{RepLinAchr_2}
\end{eqnarray}

\vspace{-0.3cm}
\noindent
\begin{eqnarray}
\left(I + M_{cx} + \ldots
+ M_{cx}^{n - 1}\right)
\cdot 
\left(r_{16}^c,\,r_{26}^c\right)^{\top}
= 
\left(0,\,0\right)^{\top}
\label{RepLinAchr_1}
\end{eqnarray}

\vspace{-0.2cm}
\noindent
must be satisfied. 
There are two possibilities regarding solutions of these equations.
Either $r_{11}^c + r_{22}^c \neq 2$, $M_{cx}^n = I$, and $r_{16}^c$ and $r_{26}^c$ are arbitrary,
or $M_{cx} = I$ and $r_{16}^c = r_{26}^c = 0$. 
In both cases the cell matrix $\,M_c\,$ can be represented in the form

\noindent
\begin{eqnarray}
M_c\,=\,D_c\, N_c\, D^{-1}_c ,
\label{TWO_C_3}
\end{eqnarray}

\noindent
where the matrix 

\noindent   
\begin{eqnarray}
N_c \,=\,
\left(
\begin{array}{cccccc}
r_{11}^c & r_{12}^c & 0        & 0        & 0 & 0 \\
r_{21}^c & r_{22}^c & 0        & 0        & 0 & 0 \\
0        & 0        & r_{33}^c & r_{34}^c & 0 & 0 \\
0        & 0        & r_{43}^c & r_{44}^c & 0 & 0 \\
0        & 0        & 0        & 0        & 1 & C \\
0        & 0        & 0        & 0        & 0 & 1
\end{array}
\right) 
\label{DispFreeMap}
\end{eqnarray}

\noindent   
is dispersion-free and the matrix $D_c $
can be expressed in the form of a Lie operator as follows

\noindent   
\begin{eqnarray}
:D_c: \,=\, \exp(:\varepsilon \, (B \, x  \,-\, A \, p_x):).
\label{TWO_C_4}
\end{eqnarray}

\noindent   
If $r_{11}^c + r_{22}^c \neq 2$, then the decomposition (\ref{TWO_C_3}) is unique,

\noindent   
\begin{eqnarray}
A\,=\,\frac{r_{16}^c - r_{52}^c}{2 - r_{11}^c - r_{22}^c}
\;\;\;
\mbox{and}
\;\;\;
B\,=\,\frac{r_{26}^c + r_{51}^c}{2 - r_{11}^c - r_{22}^c}
\label{ABC_2_1}
\end{eqnarray}

\noindent   
are the initial conditions for the periodic
(matched) cell dispersion and its derivative, and 

\noindent   
\begin{eqnarray}
C\,=\,r_{56}^c\,+\,\frac{r_{16}^c r_{51}^c + r_{26}^c r_{52}^c}{2 - r_{11}^c - r_{22}^c}.
\label{ABC_2_2}
\end{eqnarray}

\noindent   
And in the second case, when $M_{cx} = I$ and $r_{16}^c = r_{26}^c = 0$, the matrix $N_c$
is equal to the cell matrix $M_c$ and $A$ and $B$ can be chosen arbitrarily
(for example, $A = B = 0$).

\vspace{-0.2cm}
\subsection{Representation of Second-Order Aberrations\\
in the Form of a Single Lie Exponent}

Using (\ref{TWO_C_3}) the cell transfer map
can be written as

\noindent
\begin{eqnarray}
:{\cal M}_c: \,=_2\,
:D_c:^{-1}
\exp(:{\cal P}_3^c(\mbox{\boldmath $z$}):) 
\,:N_c:\, :D_c:,
\label{TWO_C2_1}
\end{eqnarray}

\noindent
where 
$\,{\cal P}_3^c(\mbox{\boldmath $z$}) =
{\cal F}_3^c(x + A \varepsilon,\, p_x + B \varepsilon, \,y,\, p_y,\, \varepsilon)$,
and for the map of the repetitive $n$-cell system ${\cal M}_{nc}$ 
we obtain after some straightforward manipulations

\noindent
\begin{eqnarray}
:{\cal M}_{nc}:\,=_2\,
\exp(:n \,{\cal S}_3(D_c^{-1}\mbox{\boldmath $z$}) \,-\, n \,C \varepsilon^2 / 2:).
\label{TWO_C2_3}
\end{eqnarray}

\noindent
In this representation the aberration function ${\cal S}_3$ is given by

\vspace{-0.2cm}

\noindent
\begin{eqnarray}
{\cal S}_3(\mbox{\boldmath $z$}) \,=\,
\frac{1}{n} \sum_{m=1}^{n-1}
{\cal P}_3^c(\hat{M}_c^m \,\hat{\mbox{\boldmath $z$}}, \,\varepsilon)
\label{FOUR_RO}
\end{eqnarray}

\vspace{-0.2cm}

\noindent
and is not an arbitrary polynomial anymore. It is 
the result of the application of the Reynolds
(averaging) operator of the cyclic group $C_n$ generated by the 
matrix $\hat{M}_c$ to the polynomial ${\cal P}_3^c$ and
for an arbitrary ${\cal P}_3^c$ is a 
polynomial which remains invariant under the group action.

As an abstract object the group $C_n$ is unique and
for all possible matrices $\hat{M}_c$ 
satisfying (\ref{TWO_C_2_HAT}) we have groups which are isomorphic each other,
but not all of them are conjugate. So that as groups of symmetries they can be
distinct and can have different number of invariant homogeneous polynomials
(remaining aberrations in (\ref{FOUR_RO})), and this depends on the choice
of the periodic cell phase advances $\mu_x^c$ and $\mu_y^c$.
For the mid-plane symmetric system the polynomial ${\cal F}_3^c$
(and therefore the polynomial ${\cal P}_3^c$)
can have as much as 18 nonzero monomials responsible for the independent 
transverse aberrations, while with the proper selection of the cell phase advances
the number of independent transverse aberrations of the $n$-cell system
can be reduced to six for $n = 2, 3$ and to two for $n \geq 4$ ~\cite{Brown}.

\vspace{-0.2cm}
\section{SECOND-ORDER ABERRATIONS OF STRAIGHT DRIFT-QUADRUPOLE CELL}

The map of the straight drift-quadrupole cell ${\cal M}_s$ 
does not have second order geometric aberrations, does not generate second order 
dispersions and the transverse motion still remains uncoupled with
the first nonlinear correction terms taken into account. Thus it can be
written as

\noindent
\begin{eqnarray}
:{\cal M}_s: \,=_2\,
\exp(:{\cal F}_3^s(\mbox{\boldmath $z$}):) :M_s:,
\label{TWO_DQ}
\end{eqnarray}

\vspace{-0.2cm}

\noindent
where

\vspace{-0.2cm}

\noindent
\begin{eqnarray}
{\cal F}_3^s(\mbox{\boldmath $z$}) 
= -\frac{\varepsilon}{2}\cdot
\left(
{\cal Q}_x(x, p_x) + {\cal Q}_y(y, p_y)
\,-\,l_s \,\frac{\varepsilon^2}{\gamma_0^2}
\right),
\label{MKJ_1}
\end{eqnarray}

\noindent
${\cal Q}_x$ and ${\cal Q}_y$ are quadratic forms,
$l_s$ is the cell length, and $\gamma_0$ is the Lorentz factor
of the reference particle.

The structure of the second-order aberrations (\ref{MKJ_1})
can be further clarified using that
for every drift-quadrupole system (which is not a pure drift space)
there exists an unique set of Twiss parameters 
(apochromatic Twiss parameters), which will be transported through
that system without first order chromatic distortions ~\cite{ApochrIPAC10}. 
Let $\,\beta_{x,y}^a$, $\,\alpha_{x,y}^a\,$ and $\,\gamma_{x,y}^a$ be these
apochromatic Twiss parameters and 

\noindent
\begin{eqnarray}
\left\{
\begin{array}{l}
I_x^a(\tau) \, = \,
\gamma_x^a(\tau)\, x^{2}  + 2 \alpha_x^a(\tau) \, x\, p_x 
+ \beta_x^a(\tau) \, p_x^{2}
\vspace{0.2cm}\\
I_y^a(\tau) \, = \,
\gamma_y^a(\tau)\, y^{2} \, + 2 \alpha_y^a(\tau) \, y\, p_y 
+ \beta_y^a(\tau) \, p_y^{2} 
\end{array}
\right.
\label{IFB_1}
\end{eqnarray}

\noindent
the corresponding Courant-Snyder quadratic forms.
Then, as it was shown in ~\cite{ApochrIPAC10},
the quadratic forms ${\cal Q}_{x,y}$ can be expressed
through these Courant-Snyder quadratic forms taken at the
cell entrance as follows 

\noindent
\begin{eqnarray}
{\cal Q}_{x, y} \,=\,
\xi_{x,y}(\beta_{x,y}^a) \cdot
I_{x,y}^a(0),
\label{IFB_2}
\end{eqnarray}

\noindent
where $\,\xi_x(\beta_x^a)\,$ and 
$\,\xi_y(\beta_y^a)\,$ are the cell chromaticities
calculated for the apochromatic Twiss parameters.

\vspace{-0.2cm}
\section{COMBINED SYSTEM AS SECOND-ORDER ACHROMAT}

We now turn our attention to the main subject of this paper.
Let us consider a system where a bend magnet block arranged 
in an achromat-like fashion is followed by a straight drift-quadrupole cell.
The formulas (\ref{TWO_C2_3}) and (\ref{TWO_DQ}) tell us that 
the map of the combined system ${\cal M}_s ({\cal M}_{nc})$
will not have transverse second-order aberrations, if and only if

\vspace{-0.2cm}

\noindent
\begin{eqnarray}
n \cdot {\cal S}_3(x - A \varepsilon,\, p_x - B \varepsilon, \,y,\, p_y,\, \varepsilon)\,-
\nonumber
\end{eqnarray}

\vspace{-0.3cm}

\noindent
\begin{eqnarray}
n \cdot {\cal S}_3(- A \varepsilon,\, - B \varepsilon, \,0,\, 0,\, \varepsilon) \,=
\nonumber
\end{eqnarray}

\vspace{-0.3cm}

\noindent
\begin{eqnarray}
\frac{\varepsilon}{2}\cdot
\left(
\xi_x(\beta_x^a) \cdot I_x^a(0)\,+\,
\xi_y(\beta_y^a) \cdot I_y^a(0)
\right).
\label{FIN_1}
\end{eqnarray}

\noindent
This equation gives the necessary and sufficient conditions
for the map of the combined system to be a second-order achromat and
will be  analyzed in more detail.
As the first step, let us rewrite the function ${\cal S}_3$ in the form

\vspace{-0.2cm}

\noindent
\begin{eqnarray}
{\cal S}_3(\mbox{\boldmath $z$}) \,=\,
\sum_{m=0}^{3} \varepsilon^m \cdot {\cal S}_{3,3-m}(\hat{\mbox{\boldmath $z$}}),
\label{FIN_2}
\end{eqnarray}

\vspace{-0.2cm}

\noindent
where each of the functions $\,{\cal S}_{3,m}\,$ is a homogeneous polynomial
of degree $m$ in the transverse variables $\,\hat{\mbox{\boldmath $z$}}$.
According to the mid-plane symmetry we have, additionally, that

\vspace{-0.1cm}

\noindent
\begin{eqnarray}
{\cal S}_{3,1}(\hat{\mbox{\boldmath $z$}}) \,=\, {\cal S}_{3,1}(x,\, p_x)
\label{FIN_2_1}
\end{eqnarray}

\vspace{-0.1cm}

\noindent
does not depend on the variables $y$ and $p_y$, and that

\vspace{-0.1cm}

\noindent
\begin{eqnarray}
{\cal S}_{3,2}(\hat{\mbox{\boldmath $z$}}) \,=\, 
{\cal S}_{3,2}^x(x,\, p_x)
\,+\,
{\cal S}_{3,2}^y(y, \,p_y).
\label{FIN_2_2}
\end{eqnarray}

\vspace{-0.1cm}

\noindent
Substituting (\ref{FIN_2}) into (\ref{FIN_1}) we obtain 
that the equation (\ref{FIN_1}) is equivalent to the following system of four equations:
Condition for the absence of geometric aberrations 

\vspace{-0.1cm}

\noindent
\begin{eqnarray}
{\cal S}_{3,3}(\hat{\mbox{\boldmath $z$}}) = 0,
\label{FIN_3}
\end{eqnarray}

\vspace{-0.1cm}

\noindent
two conditions for the absence of chromatic focusing terms 

\vspace{-0.1cm}

\noindent
\begin{eqnarray}
2 n \cdot {\cal S}_{3,2}^x(x,\,p_x) \,=\, 
\xi_x(\beta_x^a) \cdot I_x^a(0),
\label{FIN_4_1}
\end{eqnarray}

\vspace{-0.2cm}

\noindent
\begin{eqnarray}
2 n \cdot {\cal S}_{3,2}^y(y, \,p_y) \,=\, 
\xi_y(\beta_y^a) \cdot I_y^a(0),
\label{FIN_4_2}
\end{eqnarray}

\vspace{-0.1cm}

\noindent
and condition for the absence of second-order dispersions 

\vspace{-0.1cm}

\noindent
\begin{eqnarray}
n \cdot {\cal S}_{3,1}(x,p_x) = 
\xi_x(\beta_x^a) \cdot
\left(
\left[
\gamma_x^a(0) A + \alpha_x^a(0) B
\right] \cdot x +
\right.
\nonumber
\end{eqnarray}

\vspace{-0.3cm}

\noindent
\begin{eqnarray} 
\left.
\left[
\alpha_x^a(0) \,A \,+\, \beta_x^a(0)\, B
\right] \cdot p_x
\right).
\label{FIN_5}
\end{eqnarray}

\vspace{-0.1cm}

The first important observation from
the equations (\ref{FIN_3})-(\ref{FIN_5}) is that 
the functions in their left hand sides 
are invariants of the group generated by the matrix $\hat{M}_c$, 
and therefore so must be the functions in the right hand sides.
Otherwise these equations cannot be satisfied whatever number 
of sextupole families we will use in the first dispersive 
part of the system. 

If the matrix $M_{cx}$ is equal to the identity matrix, then for $n=2$ the function
${\cal S}_3$ has the same 18 independent transverse aberrations as
the function ${\cal F}_3^c$ and for $n \geq 3$ this number is reduced to 12.
That is not much and, if we want to use
automatic cancellation of some aberrations in the achromat-like part of the system
efficiently, we have to assume that $M_{cx} \neq I$.
But from this assumption it follows that linear in $x$ and $p_x$ functions
can not be invariants (i.e. ${\cal S}_{3,1}(x,p_x) = 0$) and thus the right 
hand side of the equation (\ref{FIN_5}) must be equal to zero.
There is only one possibility to satisfy this condition, namely one has to make
the basic cell of the achromat-like part of our system to be free from the linear
dispersions, i.e. $r_{16}^c$ and $r_{26}^c$ must be equal to zero.

The equations (\ref{FIN_4_1}) and (\ref{FIN_4_2}) are similar to each other
and we consider only the first.
The way of satisfying the equation (\ref{FIN_4_1}) depends on the choice of the
horizontal phase advance. If this phase advance is in the second
order resonance (i.e. if $\mu_x^c$ is multiple of $\pi$), then monomials
$x^2$, $x p_x$ and $p_x^2$ are invariants and the equation (\ref{FIN_4_1})
can be solved with the three sextupole families properly arranged in the
dispersive regions of the achromat-like part of the system.
If not, then there is only one functionally independent invariant
of the group
generated by the matrix $\hat{M}_c$ which is quadratic in $x$, $p_x$,
and this invariant can be chosen equal to the
Courant-Snyder quadratic form corresponding to the periodic Twiss parameters
of the matrix $M_{cx}$. So in this
situation the equation (\ref{FIN_4_1}) can be satisfied 
with one sextupole family, but
if, and only if, the periodic 
Twiss parameters of the matrix $M_{cx}$ coincide with the horizontal apochromatic 
Twiss parameters of the straight drift-quadrupole cell.

And finally, the equation (\ref{FIN_3}) can be satisfied either by sextupoles
or one can select such phase advances $\mu_{x,y}^c$ that none of the combinations
$3 \mu_x^c$ and $\mu_x^c \pm 2 \mu_y^c$ is multiple of $2 \pi$  and use 
automatic cancellation, which excludes the case $n = 3$ 
from considerations ~\cite{Brown}.   

\section{SUMMARY}

In this paper we have considered a mid-plane symmetric magnetostatic
beamline where a bend magnet $n$-cell repetitive system
with the overall linear transfer matrix equal to the identity matrix
is followed by a straight drift-quadrupole block 
(which is not a pure drift space)
and presented the necessary and sufficient conditions for this beamline
to be a second-order achromat in the form of the four equations 
(\ref{FIN_3})-(\ref{FIN_5}). Besides that, we have shown that
these equations can be satisfied using only six, four or two sextupole families.
In doing so one has to select such periodic cell phase advances 
$\mu_{x,y}^c$ that none of the combinations
$3 \mu_x^c$ and $\mu_x^c \pm 2 \mu_y^c$ is multiple of $2 \pi$ 
(which excludes from considerations the case $n = 3$) and then decide,
for each transverse plane separately, if the phase advance for this plane
($\mu_x^c$ or $\mu_y^c$) will be multiple of $\pi$ or not.
If it will be multiple of $\pi$, then one simply uses three sextupole families 
for this plane, and if not, then one sextupole family is sufficient, but one has 
to make the periodic Twiss parameters of the matrix $M_c$ for this plane
to coincide with the corresponding apochromatic 
Twiss parameters of the straight drift-quadrupole cell.
And in all cases the linear cell transport matrix $M_c$ must be free from 
dispersions.

For completeness, let us note that if the attached drift-quadrupole block
is a drift space with ${\cal Q}_{x,y} = -l_s p_{x,y}^2$, 
then the solution with the minimal number of sextupole
families requires    
$\mu_x^c$ be odd multiple of $\pi$ and 
$\mu_y^c$ be multiple of $\pi$ (from this it follows that $n = 2$), $B$ in (\ref{ABC_2_1}) equal to zero,
and six sextupole families.

\end{document}